# Towards Querying in Decentralized Environments with Privacy-Preserving Aggregation

Ruben Taelman[1], Simon Steyskal[2], Sabrina Kirrane[3]

[1]IDLab, Ghent University – imec, Belgium, ruben.taelman@ugent.be
[2]Siemens AG Austria, Austria, simon.steyskal@siemens.com
[3]Vienna University of Economics and Business, Austria, sabrina.kirrane@wu.ac.at

**Abstract.** The Web is a ubiquitous economic, educational, and collaborative space. However, it also serves as a haven for personal information harvesting. Existing decentralised Web-based ecosystems, such as Solid, aim to combat personal data exploitation on the Web by enabling individuals to manage their data in the personal data store of their choice. Since personal data in these decentralised ecosystems are distributed across many sources, there is a need for techniques to support efficient privacy-preserving query execution over personal data stores. Towards this end, in this position paper we present a framework for efficient privacy preserving federated querying, and highlight open research challenges and opportunities. The overarching goal being to provide a means to position future research into privacy-preserving querying within decentralised environments.

## 1. Introduction

The Web was originally envisaged as a free, non-discriminatory decentralised information space built upon standards and technical specifications. Although the Web brings major benefits as an economic, educational, and collaborative space, it also serves as a means for personal information harvesting. One potential solution to existing personal data harvesting practices is to enable individuals to take more control over who has access to their data in the form of personal data stores. One of the leading efforts in this space is Solid [1], which is a decentralised Web based ecosystem that gives people more control over their data by enabling everyone to have a personal *data pod*, and by providing app developers with the infrastructure needed to develop applications that work over distributed data sources. By decoupling data from applications, individuals are afforded more control over how their personal data are processed. In such a setting, the number of pods that need to be queried could potentially become very large, for instance in the case of a large social network.
Existing research into data aggregators [2, 3] could be used to improve query performance by reducing the number of sources that need to be consulted for any given query. However, one of the major limitations of current aggregation techniques, is the fact that they assume that all data is public, which is not realistic in many scenarios (e.g., in a social network context individuals may wish to share a limited amount of data with acquaintances and more data with friends). As such, query execution and aggregation techniques need to be extended to cater for different access policies. Concretely, data providers need to be able to associate access policies with data, and query engines need to be able to authenticate themselves to sources such that only authorised data is returned. Furthermore, since aggregators may be untrusted third-parties, there is a need for privacy-preserving aggregation techniques, such that malicious aggregators are not able to access unauthorised data, but authorised query engines are still able to exploit indexing.

In this position paper, we propose a high-level framework that can be used to: (i) optimise querying through privacy-preserving aggregation; and (ii) enable federated querying with access control. In addition, we present an instantiation of the proposed framework, and discuss open research challenges and opportunities.

The remainder of this article is structured as follows: In Section 2 we present the necessary background and related work. In Section 3 we introduce a motivating use case scenario. Following on from this our framework is introduced in Section 4 and the challenges and opportunities of possible instantiations are presented in Section 5. Finally we conclude the article and discuss future work in Section 6.

## 2. Background and Related Work

We start by presenting background and related work on the Solid platform, federated query processing, approximate membership functions, and access control.

**Social Linked Data.** Solid [1] is a *decentralised Web-based ecosystem* that decouples *data* from *applications*. With Solid, everyone has their own personal *data pod*, in which any kind of data can be stored. Concretely, Solid makes use of a collection of Web standards and technical specifications, including the Resource Description Framework (RDF) [4], the Linked Data stack [5], the Linked Data Platform (LDP) [6], Linked Data Notifications (LDN) [7], WebID [8], and Web Access Control (WAC) [9]. The RDF data model together with the Linked Data principles are used to give data a universal meaning, and to allow data to be linked across multiple data pods. Solid data pods implement the LDP specification which caters for RDF read-write operations via RESTful Web Application Programming Interfaces (APIs). The LDN specification is used to enable pods to communicate with each other. Using WebID, everyone has a personal online identifier that they can use to *authenticate* against a data pod. While, in turn WAC is used to specify rules that determine if agents and applications are *authorised* to read, write, append, or control RDF files. The framework described in this paper discusses how Solid could be extended to enable efficient privacy-preserving federated query evaluation over many Solid data pods.

**Federated Query Processing.** In a truly decentralised Web, data is spread over multiple sources, which means that there is no single endpoint through which all data can be retrieved. For this, federated query processing is an active area of research in which techniques are investigated to intelligently delegate the execution of parts of a SPARQL query to specific sources. In order to enable federations over many sources to scale more efficiently aggregation techniques whereby one or more independent *aggregators* continuously *crawl* sources, and maintain *data summaries* [2, 3], could be used to reduce the number of sources that need to be consulted. Query engines could use these summaries as an index structure that enables them to identify the sources that are needed to answer specific queries, which reduces the range of sources that need to be queried. In the context of this work, we extend existing query optimisation approaches by introducing an framework for efficient privacy-preserving federated query execution.

**Approximate Membership Functions.** Approximate Membership Functions (AMFs) are probabilistic data structures used to efficiently determine whether or not elements are part of a collection. Given that AMFs are probabilistic, they may produce false positives, but they always produce true negatives. Since AMFs are typically much smaller than a full dataset, they are a valuable method for pre-filtering when querying. *Bloom filters* [10] are one example of an AMF technique. A Bloom filter consists of a bitmap, and a predetermined set of hash functions. AMFs have been used in various of RDF querying scenarios, such as reducing the number of expensive I/O operations [11] during triple pattern query evaluation, improving the performance of join operations [12], and reducing the number of HTTP requests for Triple Pattern Fragments [13]. In the context of federated querying, the SPARQL ASK response has been enhanced with Bloom filters to share a summary of the matching results [14], which allows overlaps between different sources to be identified. Herein, we use Bloom filters to encode encrypted triple components that are available within each source, and aggregators are responsible for aggregating privacy-preserving summaries for several Solid data pods.

**Access Control.** Web Access Control (WAC) [9] is an RDF vocabulary and an access control framework, which demonstrates how together WebID and access control policies specified using the WAC vocabulary, can be used to enforce distributed access control. Villata et al. [15] and Sacco and Passant [16] extend the WAC vocabulary to cater for context based access control policies and privacy preferences respectively. Alternatively, Encryption-Based Access Control [17] involves encrypting RDF fragments (i.e. subjects, predicates, objects, graphs or some combination thereof) with an encryption key, such that only those that have the key are permitted to access the data, thus serving as both an authentication and an authorisation mechanism. Existing proposals involve using symmetric encryption [18], public-key encryption [19], or functional encryption [17] to generate RDF ciphers. In this paper, encryption mechanisms are used to create privacy-preserving aggregations, whereas access control policies are used to restrict access to data at query time.

## 3. Motivating Use Case Scenario

Following the Solid design principles, in the personalised address book use case scenario used to guide our work, address books are merely lists of WebIDs, and the actual contact details are stored in the respective contacts' pod. To keep this use case simple, we assume an address book of Alice that contains two contacts: Bob and Carol. In practise, such an address book could contain many more contacts. Alice has chosen to make this address book public, so that everyone is able to see everyone she knows, albeit without necessarily having access to everyone's private contact details as these are controlled via separate access control policies. We also consider Dave as a fourth person that has no relationship with anyone else.

For the sake of simplicity, we consider three hierarchical subject access groups per pod, where the members of each group can be configured for each pod: $S_E$ : *Everyone (without authentication)*, $S_A$ : *Acquaintances* ($S_A \subseteq S_E$), and $S_F$ : *Friends* ($S_F \subseteq S_A$). Assertions, of the following form, could be used to indicate that Carol considers Alice to be an acquaintance, and Bob considers Alice as a friend:

`<https://alice.pods.org/profile#me>` $\in S_A \subseteq S_E$

`<https://alice.pods.org/profile#me>` $\in S_F \subseteq S_A \subseteq S_E$

The data stored in Alices, Bobs and Carols pods, and the various access control rules are depicted in Fig. 1.

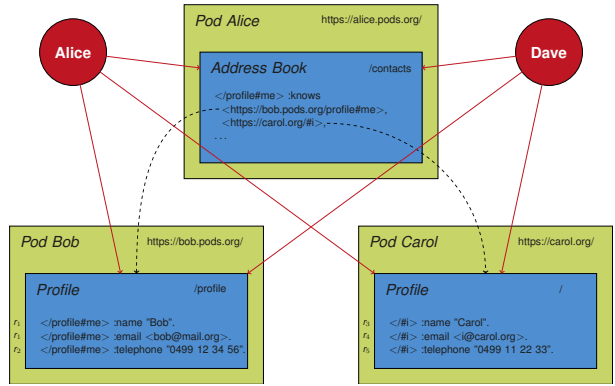

**Fig. 1:** An overview of the proposed personalised address book use case scenario.

Alice uses the `/contacts` file in her pod to list everyone that she knows using Web-IDs that point to the profiles of the respective people. The profiles of Bob and Carol both contain their name, email and telephone number, which are readable for select people. Access control rules in the form of <*subject, access rights, resource*> tuples can be used by Bob and Carol to restrict access to data stored in their respective pods: Bob is quite liberal, and allows everyone ($S_E$) to read both his name and email, however his telephone number can only be accessed by friends:

$r_1 = \langle \{s \mid s \in S_E\}, read, \{q \mid q \in Profile_B \land q.predicate \in \{:name,:email\}\} \rangle$

$r_2 = \langle \{s \mid s \in S_F\}, read, \{q \mid q \in Profile_B \land q.predicate \in \{:telephone\}\} \rangle$

Carol only allows her name to be read by the public, her email can be read by acquaintance, however her telephone number can only be accessed by friends:

$r_3 = \langle \{s \mid s \in S_E\}, read, \{q \mid q \in Profile_C \land q.predicate \in \{:name\}\} \rangle$

$r_4 = \langle \{s \mid s \in S_A\}, read, \{q \mid q \in Profile_C \land q.predicate \in \{:email\}\} \rangle$

$r_5 = \langle \{s \mid s \in S_F\}, read, \{q \mid q \in Profile_C \land q.predicate \in \{:telephone\}\} \rangle$

## 4. Efficient Privacy-Preserving Federated Querying

The goal of the proposed efficient privacy-preserving federated query execution framework, depicted in Fig. 2, is to provide a high level overview of the components needed to support privacy-preserving querying within decentralised environments.

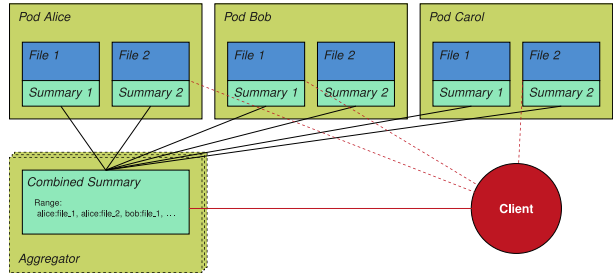

**Fig. 2:** Our efficient privacy-preserving federated query execution framework.

### 4.1. Using Data Summaries for Efficient Querying

Based on the use case scenario presented in Section 3, we assume that many data pods exist, each potentially containing multiple privacy-constrained files and, clients need to authenticate themselves to the respective data pod servers. Depending on each file's access control policy, the client may be authorised to read the full file contents, parts of it, or not at all. Since realistic decentralised environments could easily contain hundreds or thousands of files, it would be inefficient for the client to query each of them. For this reason, we make make use of the *data summaries* [2] concept in order to reduce the number of sources that need to be queried by the client. We assume that each data pod exposes a data summary for each separate file, which is subsequently aggregated by third-party aggregators, as depicted in Fig. 2. The figure provides an overview of a privacy-preserving federation with six access restricted sources and privacy-preserving summaries, and a third-party aggregator that combines these summaries in a privacy-preserving manner, together with a list of all sources it summarises. Client-side query engines can use this combined summary to derive which sources are relevant for any given query. Since files may contain private data, these data summaries must be *privacy-preserving*, i.e., they must not allow access restricted data to be leaked to unauthorised individuals. In the proposed framework, access policies are represented as access keys that are taken into account by the summary generation algorithm. Pods could generate these summaries lazily on demand, periodically or upon file changes. Following the approach from Vander Sande et al. [13], each summary consists of 4 parts, corresponding to the 4 components in RDF quads (subjects, predicates, objects and graphs). Using the summaries of these files, third-party aggregators can create *combined summaries*. Since the separate summaries are expected to be privacy-preserving, the combined summaries will also be privacy-preserving, which means that third-party aggregators need not necessarily be trusted parties. In addition to exposing the combined summary, an aggregator also needs to maintain and expose the list of sources it aggregates over, such that clients know which pods could potentially contribute query results. Although in our example we consider one aggregator, in practise multiple aggregators can exist with different source ranges. A client-side query engine can make use of the combined summary provided by the aggregator to perform source selection before query execution, i.e., reduce the number of sources to be queried. Thus the combined summaries serve to determine the pods that contain *relevant* and *accessible* data. While, the pods take care of the access control enforcement at query time, by taking into account permissions specified in terms of authorisation rules.

### 4.2. Technical Requirements

The main technical requirements are derived from the fact that our architecture needs to support efficient privacy-preserving query execution over personal data that is distributed across many sources.

- **No data leaking.** Access restricted data must not be available to those who are not authorised to access it.
- **Privacy-preserving summary creation.** It must be possible to add values to summaries by access key and file URI.
- **Summary combinations.** It must be possible to combine two summaries, where the combined summary is identical to a summary where all of the entries were added directly.
- **Authorised membership checking.** Probabilistic membership checking must be possible for a given value, access key and file URI. False positives are allowed, but true negatives are required.
- **Query Execution with Access control.** It must be possible for the pod to limit query results based on a set of access policies.

### 4.3. Core Functions of the Framework

We also propose a set of abstract algorithms that are needed in order to realise the proposed privacy-preserving federation framework. The proposed abstraction is benefical as each algorithm could be implemented in a variety of ways.

**Access Key Creation Algorithm.** As a prerequisite for encoding access into summaries, the first step is to create a map of access keys to quads based on existing access policies, using the algorithm outlined in Listing 1. Here we assume that pod owners already have a set of access control policies that govern access to quads stored in theirs pods. Although there is a many to many mapping between quads and policies, there is a one to one mapping between access policies that are used for policy enforcement at query time, and access keys that are used to create privacy-preserving summaries that are needed to optimise federated querying.

```
FUNCTION CreateAccessKeys(Q, P)
  INPUT:
    Q: set of quads, P: set of policies
  OUTPUT:
    QPK: hashmap of quads to policies and keys
QPK = new Map()
FOREACH q in Q
  FOREACH p in P
    k = GenerateKey(q,p)
    QPK = AddKey(QPK, q, p, k)
RETURN QPK
```

**Listing 1:** Algorithm for generating keys for quads based on existing access policies.

**Summary Creation Algorithm.** In the proposed framework, data pods expose a separate summary for each file, and aggregators create combined summaries using these separate summaries; and maintain a list of all source URIs that they aggregate over. We assume that pods expose summaries that are created according to the algorithm presented in Listing 2. In this algorithm, a file summary is created for each quad component, where we iterate over all the file's quads, and the access key that are applicable for each quad. For each of these combinations, we add the quad component to the summary, for the given key and file source URI. The `SummaryInitialize` and `SummaryAdd` functions that are used in the algorithm depend on the type of summary that is being used. A high-level example of this summarisation algorithm can be seen in Fig. 3.

```
FUNCTION CreatePrivacyPreservingSummary(Q, u, QPK)
  INPUT:
    Q: set of quads, u: URI of the file, QPK: hashmap relating quads to policies and keys
  OUTPUT:
    Σ: summary containing: Σ.subject, Σ.predicate, Σ.object, Σ.graph
FOREACH c in [subject, predicate, object, graph]
  Σ.c = SummaryInitialize()
FOREACH q in Q
  k = QPK(q).k
  FOREACH c in [subject, predicate, object, graph]
    Σ.c = SummaryAdd(Σ.c, q.c, k, u)
RETURN Σ
```

**Listing 2:** Algorithm for creating a summary over a file within a data pod.

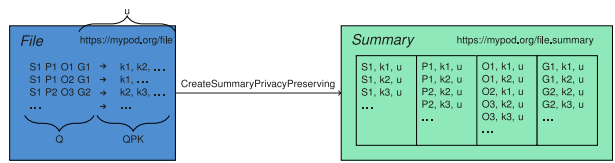

**Fig. 3:** Privacy-preserving summarisation of all RDF quads within a file.

**Summary Combination Algorithm.** Based on the resulting file summaries, the aggregator can create a combined summary using the algorithm from Listing 3. As before, the `SummaryInitialize` and `SummaryCombine` functions that are used in these algorithms depend on the type of summary that is being used. shows a high-level example of how this aggregation can happen in practise. It is worth noting that both summaries and combined summaries require some bookkeeping. Each file summary must remain up-to-date with respect to the file's contents. This could be done by either immediately invalidating the summary upon file changes, or by periodically regenerating the summary. The combined summary requires similar actions to avoid going stale. This can be achieved through immediate notifications from the pod to the aggregator upon file changes, or the aggregator can periodically scan the files or its summaries for changes.

```
FUNCTION CreateAggregatedSummary(U)
  INPUT:
    U: set of sources
  OUTPUT:
    Σ: combined summary containing: Σ.subject, Σ.predicate, Σ.object, Σ.graph
FOREACH c in [subject, predicate, object, graph]
  Σ.c = SummaryInitialize()
FOREACH u in U
  Σ' = get summaries from u
  FOREACH c in [subject, predicate, object, graph]
    Σ.c = SummaryCombine(Σ.c, Σ'.c)
RETURN Σ
```

**Listing 3:** Algorithm for creating a combined summary over a set of sources.

**Client-side Source Selection Algorithm.** Assuming we have an aggregator exposing a summary over a set of sources, we introduce the algorithm in Listing 4 where a client-side query engine can make use of an aggregator's summary to reduce the number of sources the client should query over, i.e., to perform *source selection*. In this case, we only consider quad pattern queries, because they form the foundation of more expressive SPARQL queries, and query engines typically decompose SPARQL query into several smaller quad pattern queries through a query planner [20]. As input, our algorithm assumes a quad pattern query, the list of access keys provided by the user, and the summary and list of sources it obtained from an aggregator. Based on these inputs, the client will iterate over all non-variable quad components and all available keys. For each combination, it will first do a pre-filtering step before locally iterating over all sources. It will check whether or not the quad component value is present in the summary for the current key and quad component, with source URI set to ε to match with all sources. If it is not present, then we return an empty array, as none of the sources will contain the given component value. If it is present, some of the sources *may* contain the component value, because we consider summaries as being probabilistic. After that, we iterate over each source URI, and check its presence in the summary of the current quad component, combined with the component value and key. When a true negative is found for a source, this source is removed from the list of sources. Finally, the remaining list of sources is returned, which can be used by the query engine to execute the quad pattern query over. In this algorithm, the `SummaryContains` also depends on the type of summary that is being used.

```
FUNCTION SelectSources(q, K, Σ, U)
  INPUT:
    q: quad pattern query, K: access keys, Σ: summary containing: Σ.subject, Σ.predicate, Σ.object, Σ.graph,
  OUTPUT:
    U': list of selected sources
U' = []
FOREACH c in [subject, predicate, object, graph]
  IF q.c not variable
    FOREACH k in K
      IF ! SummaryContains(Σ.c, q.c, k, ε)
        RETURN []
      FOREACH u in U
        IF ! SummaryContains(Σ.c, q.c, k, u)
          add u to U'
RETURN U'
```

**Listing 4:** Client-side algorithm for selecting query-relevant sources.

**Client-side Query Execution Algorithm.** Once the client has obtained the list of sources that it needs to query for a given quad pattern, the next step is to execute the query against each source. The client uses the sources returned by the aggregator to execute queries against the various pods using the algorithm outlined in Listing 5. The algorithm takes as input a client identification (e.g., WebID), a quad pattern query, and the set of sources returned by the source selection algorithm. Individual queries are executed against each of the sources and the aggregated results are returned to the client.

```
FUNCTION QuerySources(i, q, U)
  INPUT:
    i: client identification, q: quad pattern query, U: list of sources
  OUTPUT:
    R: query results
R = {}
FOREACH u in U
  R = R U ExecuteQuery(i, q, u)
RETURN R
```

**Listing 5:** Client-side algorithm for querying query-relevant sources.

**Server-side Query Execution Algorithm.** On receipt of a query the server uses the algorithm outlined in Listing 6 to ensure that only authorised query results are returned to the client. In the proposed algorithm a map relating quads to policies and keys is used to identify access policies that govern a particular query. We assume that there may be multiple policies that govern a particular quad and thus envisage a simple conflict resolution strategy whereby either prohibitions override permissions or visa versa. The algorithm stops as soon as it finds a policy that permits the given query to be executed and returns the results of the query execution.

```
FUNCTION ExecuteQuery(i, q, QPK)
  INPUT:
    i: client identification, q: quad pattern query, QPK: hashmap relating quads to policies and keys
  OUTPUT:
    R: query results
R = {}
IF i not verified
  RETURN {}
p = QPK(q).p
IF ! AllowedAccess(p, i, q)
  RETURN {}
ELSE
  R = ExecuteQueryWithAccessControl(i, q, p)
RETURN R
```

**Listing 6:** Querying with access control algorithm.

### 4.4. Query Execution Over Privacy-Preserving Summaries

Fig. 4 shows an example of how our privacy preserving summaries can be used in client-side query engines. The presented high level architecture should be seen as a basis for federated querying over decentralised environments with private data, where there is a single aggregator, and all sources we want to query over are considered by the aggregator. In practise, multiple aggregators can exist, they may apply to overlapping sources, and some sources may not be aggregated at all. For these cases, extensions to this algorithm will be needed, which we consider out-of-scope for this work.

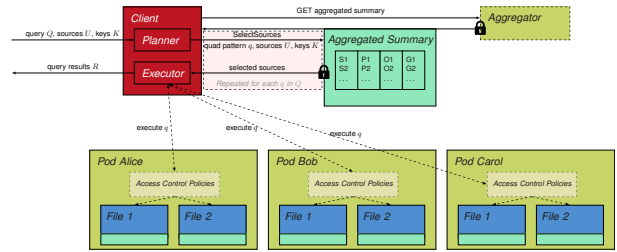

**Fig. 4:** Federated query execution using a privacy-preserving summary.

## 5. Challenges and Opportunities

When it comes to efficient privacy-preserving federated query evaluation over many Solid data pods there are several open challenges and opportunities.

**Access Policy Specification.** We assume that pod owners need to be able to specify access control policies that can be enforced both from an indexing and also a query processing perspective, taking into consideration the **no data leaking** requirement. Considering our use case scenario, in order to support privacy-preserving summaries, there is a need to generate access keys for both the *acquaintances* and the *friends* files, such that the summary generation process does not work with plain text attributes but rather cipher text. In the case where data is public by default, for instance in the case of the *everybody* file, no key is needed. We assume that there is a many to many mapping between quads and policies and a one to one mapping between access policies (enforced at query time) and access keys (used to create privacy preserving data summaries). Our initial proposal makes use of simple symmetric keys, however for more complex scenarios both attribute-based encryption and/or key derivation algorithms could be use to provide support for more complex access policies. When to comes to policy management, there is a need to ensure that (i) access keys are tightly bound to access policies, and (ii) said keys are distributed to authorised individuals (i.e. acquaintances and friends). In order to revoke access to a particular individual one would need to regenerate the keys and redistribute them to authorised individuals.

**Summary Generation and Maintenance.** The requirements for enabling federated querying in an efficient manner through privacy-preserving aggregators are mainly driven by the summarisation technology. In this context symmetric keys are used to create privacy-preserving summaries that do not leak access restricted data. We consider AMFs, such as Bloom filters, as being one possible candidate for such summaries that meet the **privacy-preserving summary creation.** and **summary combinations** requirements. The main advantage of using AMFs is that all of the performance-critical operations on summaries (adding, combining, membership checking) can happen very efficiently, as these are essentially just bitwise operations. However, it is important to highlight that certain parameters need to be configured, and that all operations must be known before they can be operationalised. For example, for Bloom filters the parameters are the number of hashes and bits. These parameters and the number of entries all impact the false positive error rate. Concretely, the parameters used to setup individual summaries need to be identical such that they can be combined by an aggregator. In a decentralised environment, it is however difficult to reach a consensus with respect to fixed parameters. This means that a parameter determination mechanism is needed for aggregators that want to combine multiple AMFs. Also, since the creation of an AMF for a file can become expensive, sources may decide to adopt different maintenance strategies.

**Source Selection.** In the proposed framework, a client-side query engine can make use of the aggregator's summary to perform source selection, in order to reduce the number of sources that are being consulted by this engine. From a source selection perspective, we address the **authorised membership checking.** requirement. As these summaries allow source selection based on quad patterns instead of full SPARQL queries, source selection can be pushed down into the query plan, which allows quad patterns in the query to be executed over a different range of sources. Furthermore, instead of applying source selection before query execution, this allows source selection to optionally happen adaptively during query execution, following the federation algorithm of Triple Pattern Fragments [21]. A hybrid approach where source selection happens both before and during query execution could be investigated. Open challenges include investigating how file-based source selection could be combined and enhanced by existing source selection methods for SPARQL endpoints, and the automatic discovery of applicable aggregators by clients.

**Query Execution with Access control.** Since file-based APIs are the basis for data retrieval on the Web as prescribed by the HTTP protocol, we assume this as a starting point for federated querying in decentralised environments. Furthermore, we consider quad pattern-based access to file sources instead of more complex SPARQL queries. This is because triple and quad patterns are the fundamental elements of SPARQL queries, and any SPARQL query can be decomposed into multiple smaller quad pattern queries. For example, client-side query engines such as Comunica [20] decompose any SPARQL query into a sequence of quad pattern queries for evaluation against heterogeneous sources, where the results of these queries are joined together locally. More complex SPARQL features such as `FILTER` and aggregates are handled client-side. Once the query engine has identified the data sources that could potentially contribute results to their query, the query engine needs to authenticate the user to the server(s) and execute the query or parts thereof. The server is responsible for enforcing access control, and executing the query or parts thereof. Here, we address the **query execution with access control** requirements. One of the key challenges with respect to access and usage Control relates to the enforcement of authorisations, i.e., policies (cf., ) which govern *who* can do *what* with *which* resources under *what* conditions. Here, we envision a mechanism that translates access policies (i.e. sets of authorisations) into constraints (e.g., data shapes like SHACL [22]) which requests and re-spective query results can then be validated against. However, the trade-off between granularity of policies (e.g., file-based, pattern-based, quad-based, …) and associated computational overhead needs to be thoroughly investigated.

## 6. Conclusions

In this paper, we propose a framework for efficent privacy-preserving querying within decentralised environments where distributed data sources are governed by one or more access control policies. The proposed framework, which is built around the notion of privacy-preserving summaries, serves as a basis for exploring and comparing alternative strategies for efficient querying with access control. As a first step, we discuss a possible instantiation of this framework which uses Bloom filters for creating privacy-preserving summaries over encrypted data, and highlight several open research challenges and opportunities. In future work, we will evaluate the use of Bloom filters for privacy-preserving federated querying both in terms of performance and privacy preservation. Additionally, we will investigate how access control policies and access keys can be managed effectively and efficiently in a decentralised Web based ecosystem such as Solid.